 \pdfoutput=1 
\documentclass{PoS}
\usepackage{amsmath}

\usepackage{graphicx}                                                                                  
\usepackage{dcolumn}                                                                                   
\usepackage{bm}                                                                                        
\usepackage{color}                                                                                     
\usepackage{booktabs}                                                                                  
\usepackage{multirow}                                                                                  
\definecolor{grey}{rgb}{0.9,0.9,0.9}                                                                   
                                                                                                       
\usepackage{ifthen}                                                                                    
\usepackage{keyval}

\title{Measurement of the Pion Polarizability at COMPASS}

\ShortTitle{Measurement of the Pion Polarizability at COMPASS}

\author{\speaker{Stefan Huber}\thanks{
This work was supported by the BMBF, the DFG Cluster of Excellence "Origin and Structure of the Universe"~(Exc~153), and the Maier-Leibnitz-Laboratorium der Universit\"at und der Technischen Universit\"at M\"unchen.
Participation at the conference was financially supported by the organizers of the conference.
}\ on behalf of the COMPASS collaboration\\
        Technische Universit\"at M\"unchen\\
        E-mail: \email{stefan.huber@cern.ch}}

\abstract{The value of the pion polarizability is predicted with high precision
by Chiral Perturbation Theory. However, the existing experimental
values are at tension with this prediction as well as among
themselves.

The COMPASS experiment at the CERN SPS accesses pion-photon reactions
via the Primakoff effect, where high-energetic pions react with the
quasi-real photon field surrounding the target nuclei. Flagship
channel is the Primakoff reaction in which a single real photon is
produced, giving access to pion Compton scattering. Using this process
the pion polarizability is extracted from the measured cross-section
shape.

End of 2009 COMPASS performed a measurement of the pion polarizability
using a nickel target. The large amount of data collected in
combination with the possibility to study systematic effects using the
analogous reaction with a muon beam, the most precise experimental
value to date was determined.}

\FullConference{XV International Conference on Hadron Spectroscopy-Hadron 2013\\
		4-8 November 2013\\
		Nara, Japan }

\begin{document}

\section{Introduction}

The \textbf{Co}mmon \textbf{M}uon and \textbf{P}roton \textbf{A}pparatus for \textbf{S}tructure and \textbf{S}prectroscopy, 
COMPASS \cite{Abbon:2007pq} is a fixed-target experiment operated at CERN's Super Proton Synchrotron. The aim is to study the structure and spectrum of light hadrons.
Being the lightest meson, the pion is of special interest as it is the fundamental particle in low-energy QCD described within the framework of chiral perturbation theory.
For non-point-like particles the electric and magnetic polarizabilities $\alpha_\pi$ and $\beta_\pi $ are fundamental quantities describing the behavior of the particles electromagnetic fields.
For these values exist precise expectations from chiral perturbation theory \cite{Gasser:2006qa}.

\section{Primakoff reactions}

Due to the short life-time of the pion it is impossible to prepare a pion target which the photons may scatter off. One possibility to study pion-photon scattering processes 
is to use highly relativistic pion beams on nuclear targets, as has been proposed by Henry Primakoff \cite{prim}. In these processes the charged pion scatters off quasi-real photons stemming from the strong electromagnetic
field surrounding heavy nuclei. The density of these quasi-real photons can be described using the Weizs\"acker-Williams approximation \cite{Pomeranshuk:1961} in which the  cross-section can be written as
\begin{equation}
	\label{eq:weiz}
	{\frac{\text{d}\sigma_{\pi\mbox{\tiny  Ni}}^{\mbox{\tiny EPA}}}
	{\text{d}s\,\text{d}Q^2\,\text{d}\Phi_{n}}} = 
	{\frac{Z^{2}\alpha_\text{em}}{\pi (s-m_{\pi}^{2})}}\ 
	F^2_{\mbox{\tiny eff}}(Q^2)\ \frac{Q^2-Q_{\min}^2}{Q^{4}}
	\ \frac{\text{d}\sigma_{\pi\gamma}}{\text{d}\Phi_{n}},
\end{equation}
where $m_\pi$ is the mass of the pion, s the center-of-mass energy of the $\pi\gamma$-system and $\alpha_\text{em}$ the fine-structure constant with a value of about $1/137$.
$Q_\text{min}$, the minimal four-momentum transfer is a quantity determined by the reaction kinematics and given by the formula
\begin{equation}
	Q_\text{min}=\frac{s-m_\pi^2}{2p_\text{beam}}.
\end{equation}
F$_\text{eff}$ is the nuclear form factor and $\sigma_{\pi\text{Ni}}$ is proportional to the squared nuclear charge $Z^2$. For the extraction of the pion polarizability we consider the case of pion Compton scattering 
$\pi^-\gamma\rightarrow\pi^-\gamma$ which is related by Eq.\ref{eq:weiz} to the process $\pi^-\text{Ni}\rightarrow\pi^-\gamma\text{\,Ni}$. The polarizabilities enter the $\pi\gamma$ cross-section as given by.

\begin{equation}
	\frac{\text{d}\sigma_{\pi\gamma}}{\text{d}\Omega} = 
	\left(\frac{\text{d}\sigma_{\pi\gamma}}{\text{d}\Omega}\right)_{Born}
	-\frac{\alpha_{em}\, m_\pi^3(s-m_\pi^2)^2}{4s\,(s\, z_+ + m_\pi^2\, z_-)}
	\left(z_-^2(\alpha_\pi-\beta_\pi)
	+\frac{s^2}{m_\pi^4}z_+^2(\alpha_\pi+\beta_\pi)\right),
	\label{eq:xsec}
\end{equation}
where $z_\pm$ is given by $1\pm\cos\theta_{cm}$ respectively and the $\theta_{cm}$ is the scattering angle of the pion in the center-of-mass system.

\section{Measurement at COMPASS}

\begin{figure}
	\begin{center}
		\includegraphics[width=0.46\textwidth]{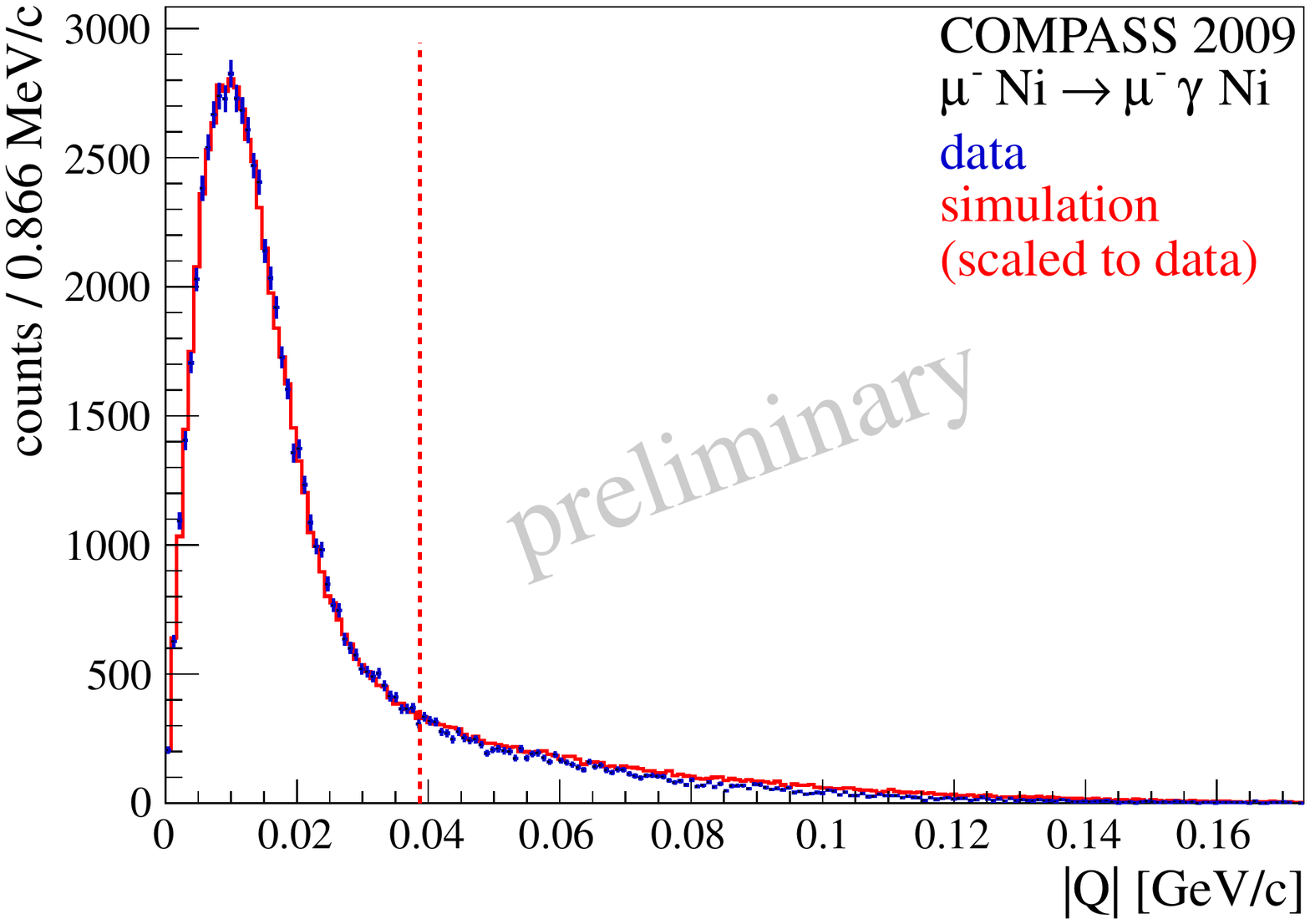}
		\hfill
		\includegraphics[width=0.46\textwidth]{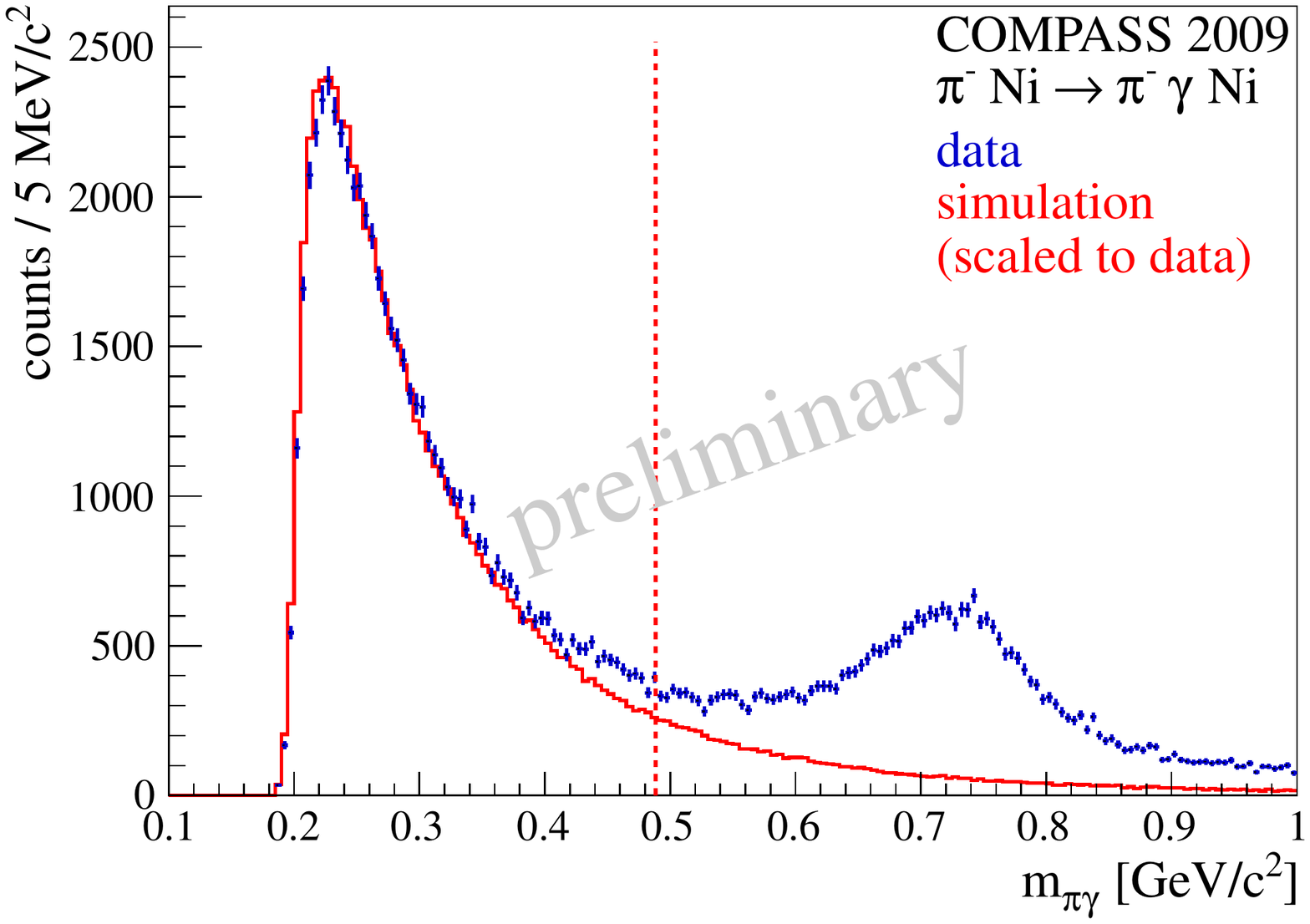}
		{\centerline{\footnotesize\sffamily\bfseries (a) \hspace{.5\textwidth} (b)}}
		{\\}
		\includegraphics[width=0.46\textwidth]{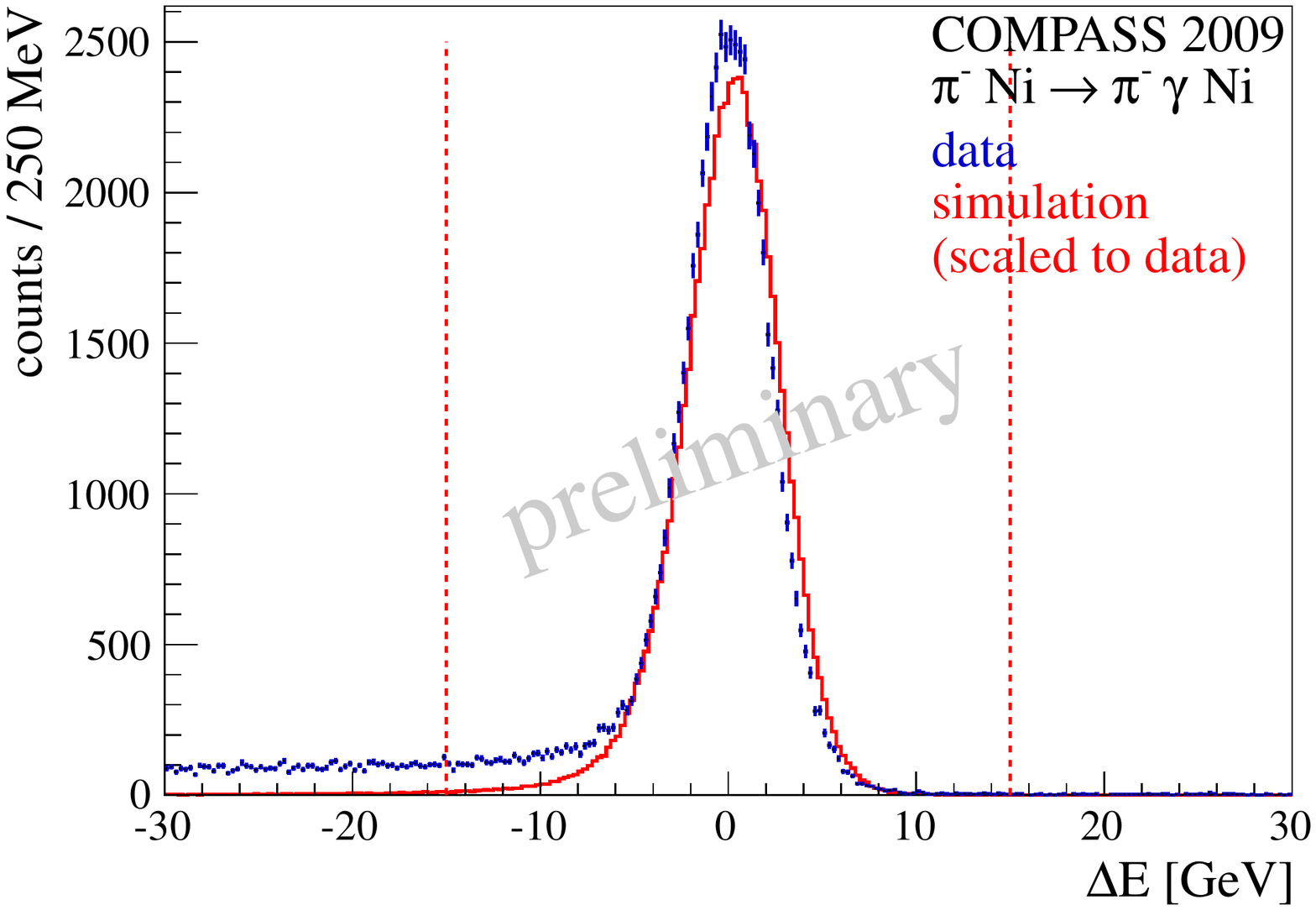}
		\hfill
		\includegraphics[width=0.46\textwidth]{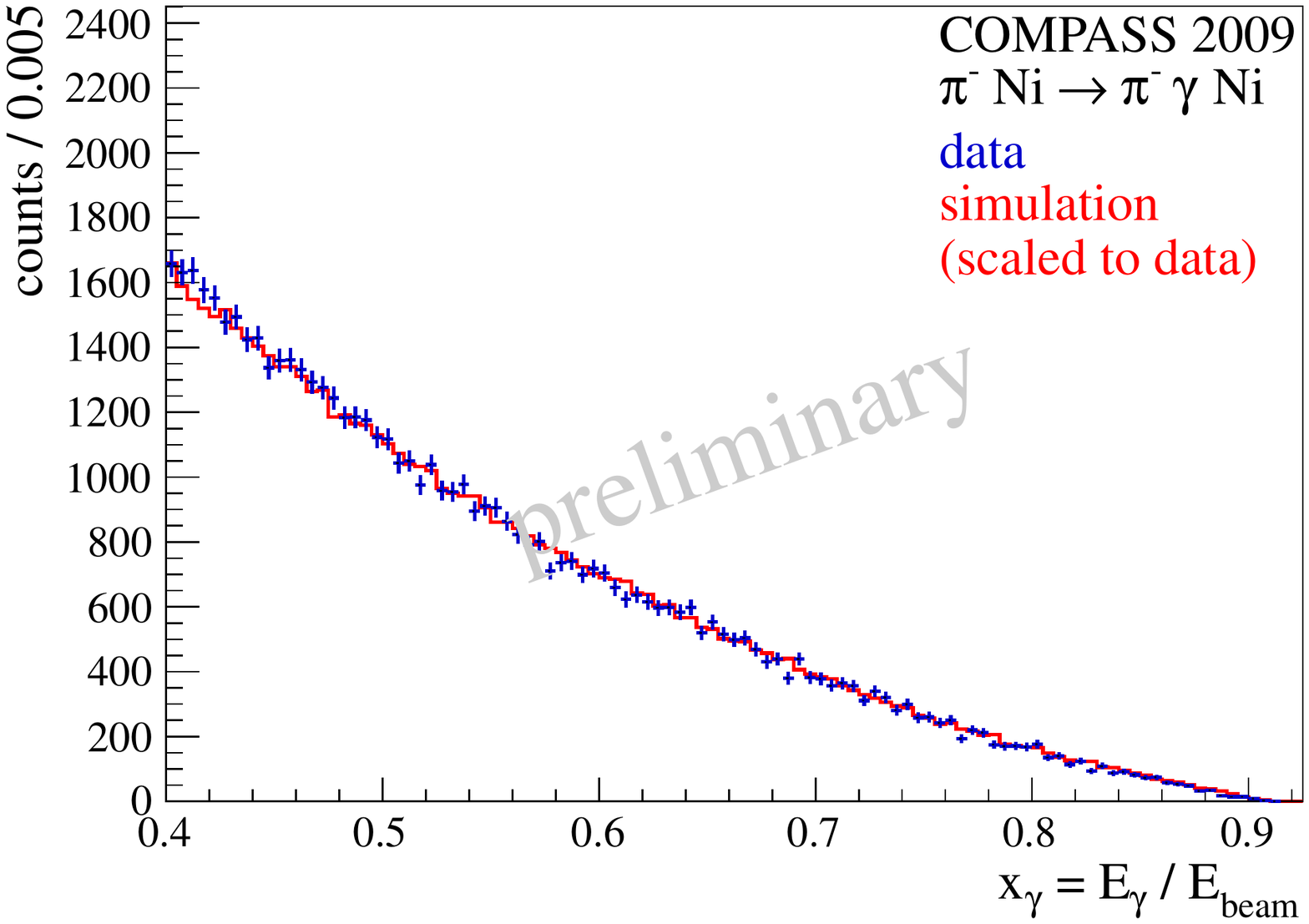}
		{\centerline{\footnotesize\sffamily\bfseries (c) \hspace{.5\textwidth} (d)}}
		{\\}
		\caption{Kinematic distributions for the reaction $\pi^-\text{Ni}\rightarrow\pi^-\gamma\text{\,Ni}$. Magnitude |Q|
		of the four-momentum transfer (a), invariant mass of the $\pi\gamma$ system (b), energy balance (c) and relative energy $x_\gamma$ of the photon (d).
		}
		\label{fig:a}
	\end{center}
\end{figure}

For this measurement COMPASS uses a beam of negatively charged hadrons consisting mainly of $\pi^-$. A kaon contamination of about 3\% is rejected using particle identification by Cherenkov counters
installed at the end of the beam line in front of the target. The pions impinge on a 3\,mm thick nickel disk. The reaction $\pi^-\text{Ni}\rightarrow\pi^-\gamma\text{\,Ni}$ is identified by selecting events with one 
outgoing charged track and one cluster in the electromagnetic calorimeter having an energy $E_\gamma$ above $80\text{\,GeV}$. Taking into account the resolution of the electromagnetic calorimeter the reaction is
required to be exclusive by cutting on the energy balance $\Delta E=\left|E_\text{beam}-E_\gamma-E_{\pi}\right|<15\,\text{GeV}$ figure \ref{fig:a}c. The remaining non-exclusive background originates mainly from processes having one $\pi^0$ in the final state, where one of the photons from the decay $\pi^0\rightarrow\gamma\gamma$ is not reconstructed. This can be seen from the remaining $\rho^-(770)$ peak in the invariant mass distribution. 

The cross-section for quasi-real photon exchange diverges as with decreasing four-momentum-transfer Q as given by Eq.~\ref{eq:weiz}. Using this property Primakoff events are selected by applying a cut on the squared four-momentum transfer of $Q^2<0.0015\,\text{(GeV/c)}^2$. This also effectively suppresses contributions from diffractive processes, where the interaction probability drops approximately linearly with $Q^2$. In order to illustrate the momentum-transfer distribution, the quantity $|Q|=\sqrt{|Q^2|}$ is shown in Fig. \ref{fig:a}a. In order to achieve the experimental resolution necessary to identify this process, special care was taken to properly align the vertex detectors as well as to properly calibrate the electromagnetic calorimeter. At very low scattering angles multiple scattering dominates. In order to select only events where the interaction vertex can be properly determined, a cut on the transverse outgoing momentum of the pion w.r.t to the beam is applied $p_T>40\,\text{MeV/c}$.

\begin{figure}
	\begin{center}
		\includegraphics[width=0.46\textwidth]{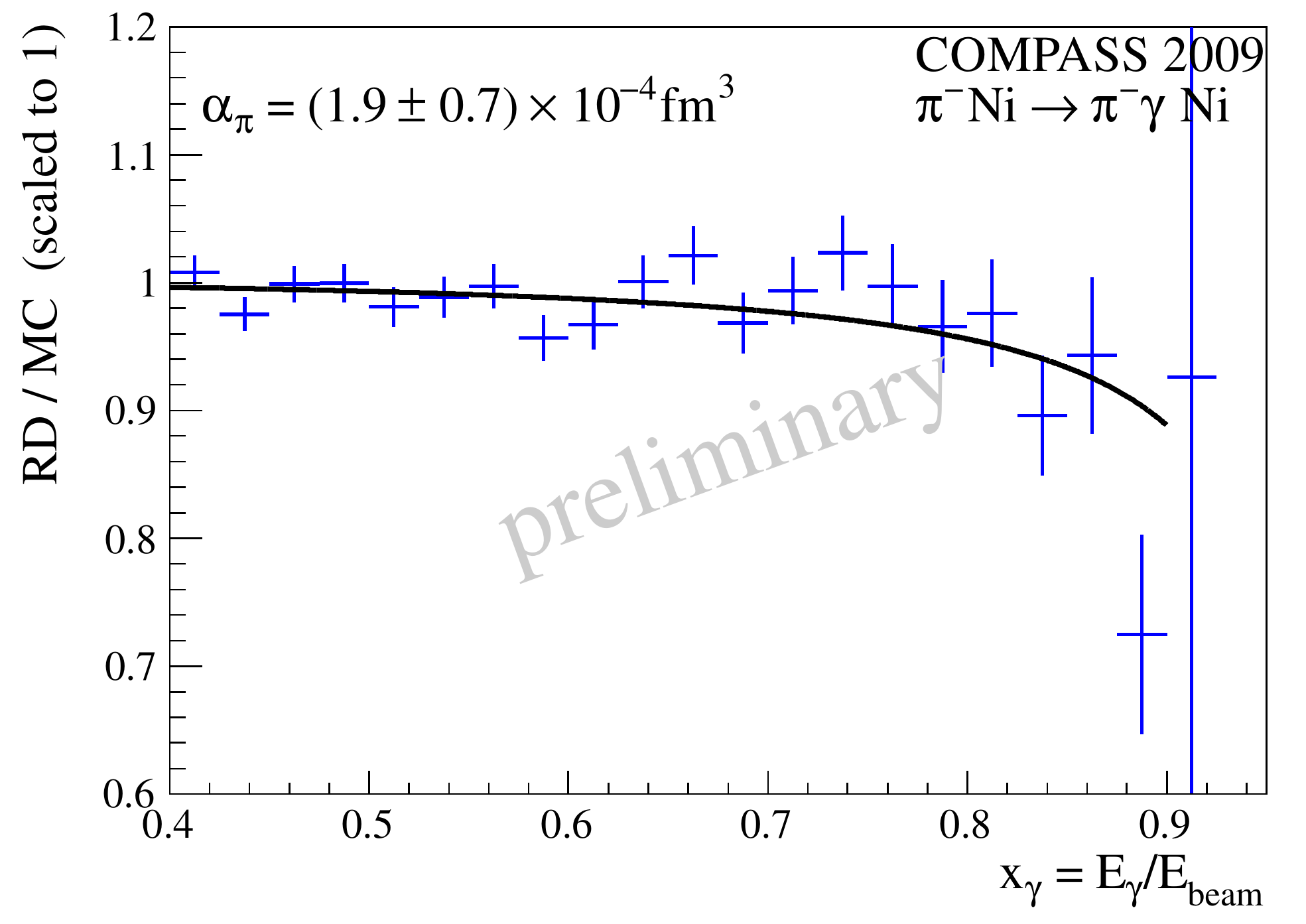}
		\hfill
		\includegraphics[width=0.46\textwidth]{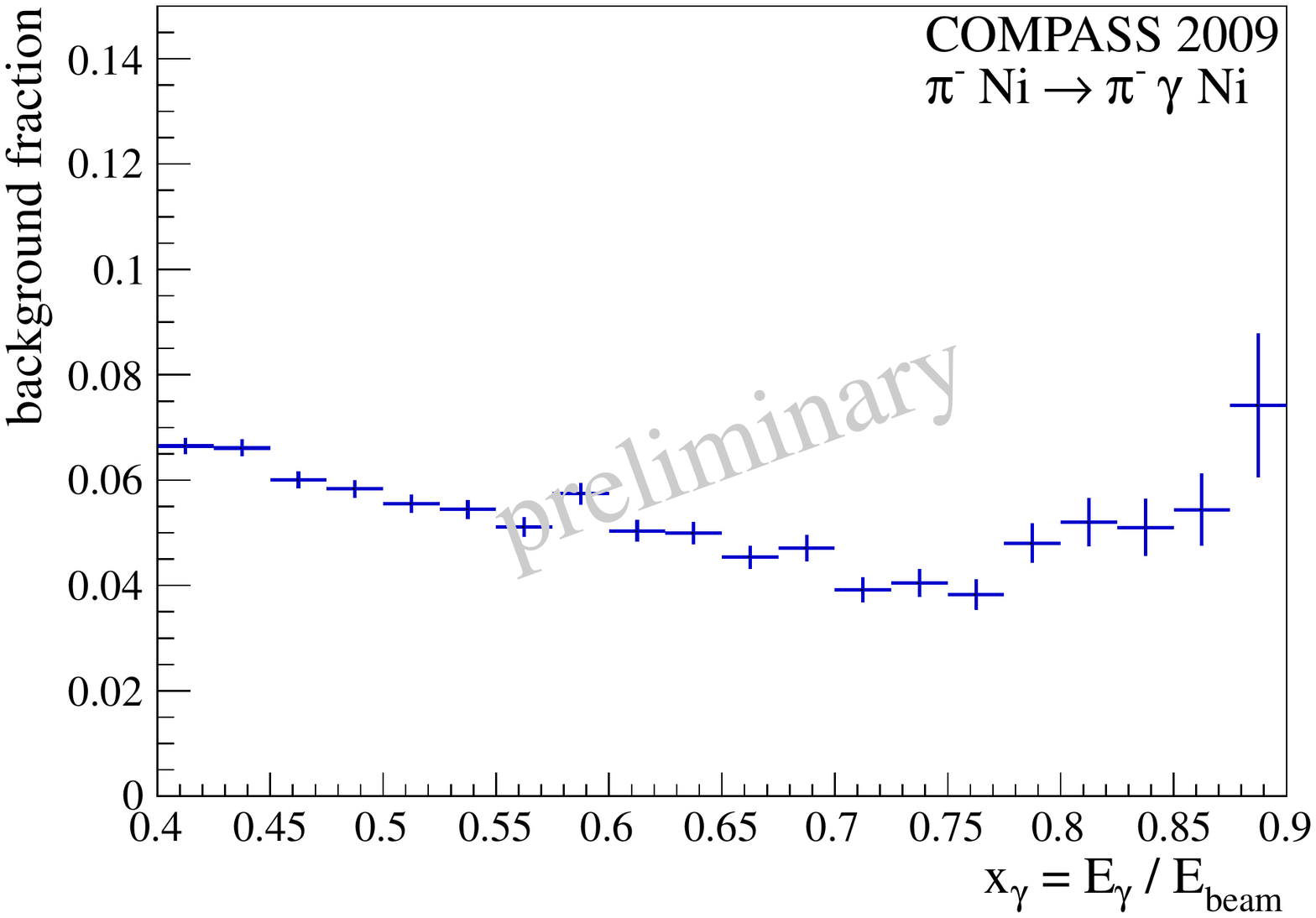}
		{\centerline{\footnotesize\sffamily\bfseries (a) \hspace{.5\textwidth} (b)}}
		\caption{Extraction of the polarizability value from the background subtracted data a) and relative background fraction b).}
		\label{fig:xratio}
	\end{center}
\end{figure}

By using the approximation $\alpha_\pi=-\beta_\pi$ the ratio of the Born cross-section, which describes a point-like particle and the cross-section taking into account the polarizability contribution 
integrated over the considered range of four-momentum transfer reduces to a simple formula depending only on the value $x_\gamma=E_\gamma/E_{beam}$. 
\begin{equation}        
	R\ =\ \left(\frac{\text{d}\!\sigma}{\text{d}\!x_\gamma}\right)
	\left/
	\left(\frac{\text{d}\!\sigma^{0}}{\text{d}\!x_\gamma}\right)
	\right.
	= 1 \ -\  \frac{3}{2}\cdot \frac{m_\pi^3}{\alpha}
	\cdot \frac{x_\gamma^2}{1-x_\gamma}\ \alpha_\pi
	=\frac{\text{dN}_\text{Born}/dx_\gamma}{\text{dN}_\text{meas.}/dx_\gamma}
	\label{eq:rxg}
\end{equation}
Applying this formula to our measurement requires the description of the cross-section for a point-like spin-0 particle. For that the $x_\gamma$-distribution is
taken from a Monte Carlo simulation. The ratio between the measured $\pi^-$-spectrum and the simulation is depicted in figure \ref{fig:xratio}. 
A polarizability value of $\alpha_\pi=(1.9\pm0.7_\text{stat})\cdot10^{-4}\text{\,fm}^3$ is extracted by fitting Eq.~\ref{eq:rxg} to this spectrum.

In order to check systematic effects on the measurement COMPASS takes advantage of the possibility to easily change from hadron to muon beam. 
For muons, which are point-like particles, the polarizability is zero and so a measured "false polarizability" value for muons gives an estimate of the 
systematic uncertainties stemming from the Monte Carlo description of the experiment. The extracted uncertainty on the polarizability according to this procedure
is $\pm0.6\cdot10^{-4}\text{\,fm}^3$.
\begin{table}[h]
	\begin{center}
			\label{tab:systematics2}
			\begin{tabular}{lr}
				\toprule
				\multirow{2}{*}{uncertainty source} & estimated magnitude\\
																									& $\text{CL}=68\,\%\quad[10^{-4}\,\text{\,fm}\rule{0pt}{7pt}^3]$\\
				\midrule
				tracking                                  & 0.6\\
				radiative corrections                     & 0.3\\
				background subtraction in $Q$             & 0.4\\
				pion-electron scattering                  & 0.2\\
				\midrule
				\midrule
				quadratic sum                             & 0.8\\
				\bottomrule
			\end{tabular}
		\end{center}
		\caption{Systematic uncertainties estimated
			on 68\,\% confidence level}
				\label{tab:tab}
	\end{table}
The systematic uncertainty from  the applied  background subtraction procedure is estimated to be of the order of $\pm0.4\cdot10^{-4} \text{\,fm}^3$. One additional contribution
of background, which is not yet taken into account, is the scattering of the pions on the electrons of the target, which then loose their energy by photon emission and are therefore missidentified.
The systematic effect due to this process is estimated as $\pm0.2\cdot10^{-4} \text{\,fm}^3$. All given systematic uncertainties are summarized in table~\ref{tab:tab}.
Adding up these contributions quadratically yields a total systematic uncertainty $\pm0.8\cdot10^{-4} \text{\,fm}^3$ which in conclusions leads to a preliminary result of the pion polarizability of
\begin{equation}
\alpha_\pi=(1.9\pm0.7_\text{stat}\pm0.8_\text{sys})\cdot10^{-4} \text{\,fm}^3.
\end{equation}
\section{Conclusion}
\begin{figure}
	\begin{center}
		\includegraphics[width=0.46\textwidth]{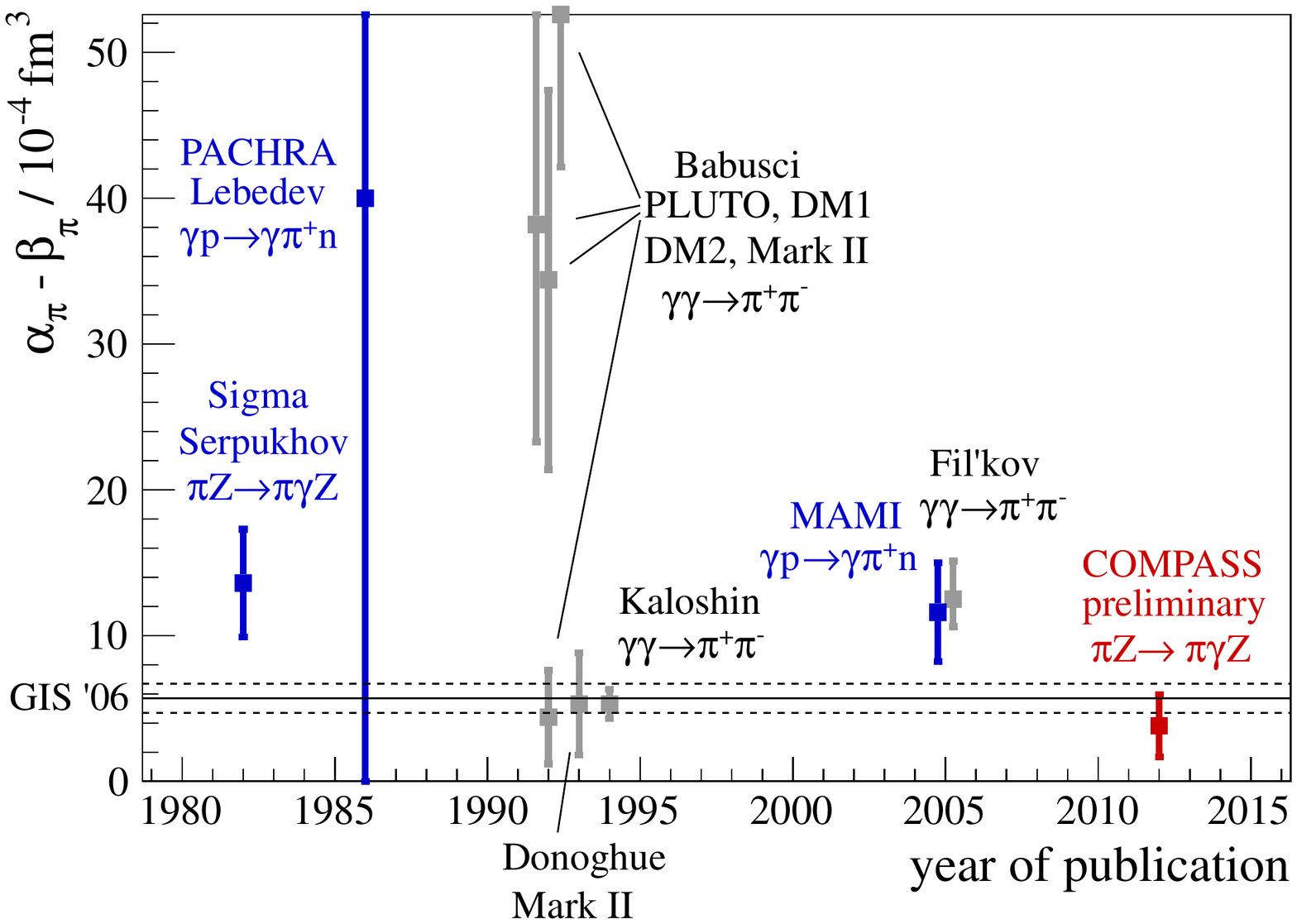}
		\hfill
		\includegraphics[width=0.46\textwidth]{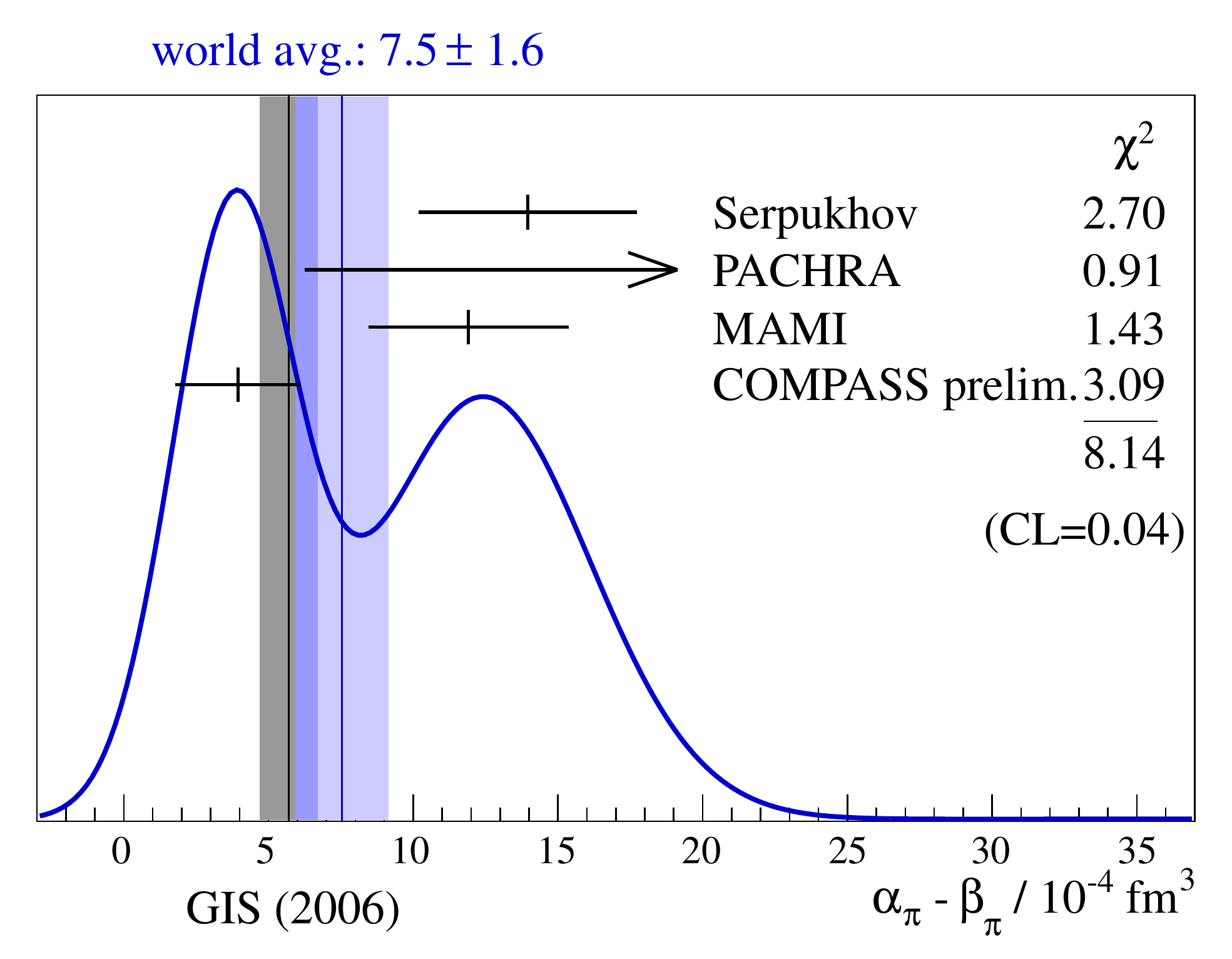}
		{\centerline{\footnotesize\sffamily\bfseries (a) \hspace{.5\textwidth} (b)}}
		\caption{Overview of the world data for the value $\alpha_\pi-\beta_\pi$ including the COMPASS data a) and 
		an ideogram for the same data as given in the PDG \cite{PDG} where the value indicated with GIS corresponds to the prediction of chiral perturbation theory \cite{Gasser:2006qa}.
		For a detailed discussion of the experimental points refer to \cite{Nagel}.
		}
		\label{fig:ideo}
	\end{center}
\end{figure}

The COMPASS result based on the Primakoff technique is in tension with previous measurements of the pion polarizability as shown in figure \ref{fig:ideo}.
In contrast it is in good agreement with the theoretical value from chiral perturbation theory. 

In order to further improve the experimental precision COMPASS has taken a bigger data set during the 2012 run. The experimental approach to this data is similar  
and will allow an independent extraction of $\alpha_\pi$ and $\beta_\pi$. Furthermore identifying the kaon contribution in the beam will allow to extract the polarizability value for kaons as well.

\end{document}